# Quantum Oblivion: A Master Key for Many Quantum Riddles

Avshalom C. Elitzur[1], Eliahu Cohen[2]

[1]*Iyar*, The Israeli Institute for Advanced Research, Rehovot, Israel.
avshalom@iyar.org.il
[2]School of Physics and Astronomy, Tel Aviv University, Tel-Aviv 6997801, Israel.
eliahuco@post.tau.ac.il

A simple quantum interaction is analyzed, where the paths of two superposed particles asymmetrically cross, while a detector set to detect an interaction between them remains silent. Despite this negative result, the particles' states leave no doubt that a peculiar interaction has occurred: One particle's momentum changes while the other's remains unaffected, in apparent violation of momentum conservation. Revisiting the foundations of the quantum measurement process offers the resolution. Prior to the macroscopic recording of no interaction, a brief Critical Interval prevails, during which the particles and the detector's pointer form a subtle entanglement which immediately dissolves. It is this self-cancellation, henceforth "Quantum Oblivion (QO)," that lies at the basis of some well-known intriguing quantum effects. Such is Interaction-Free Measurement (IFM) [1] and its more paradoxical variants, like Hardy's Paradox [2] and the Quantum Liar Paradox [3]. Even the Aharonov-Bohm (AB) effect [4] and weak measurement [5] turn out to belong to this group. We next study interventions within the Critical Interval that produce some other peculiar effects. Finally we discuss some of the conceptual issues involved. Under the time-resolution of the Critical Interval, some nonlocal phenomena turn out to be local. Momentum is conserved due to the quantum uncertainties inflicted by the particle-pointer interaction, which sets the experiment's final boundary condition.

**PACS:** 03.65.Ta, 03.65.Ud, 03.65.Xp, 11.30.-j.

**Keywords:** Quantum Oblivion, quantum measurement, IFM, AB effect, quantum Zeno effect, nonlocality.

# 1. INTRODUCTION

Momentum conservation is one of classical physics' most fundamental laws, which every translational symmetric system must obey. A basic quantum interaction, however, seems to defy it, and the resulting paradox's resolution offers novel results and insights.

This article's outline is as follows. The interaction in question is presented in Section 2. In 3 we point out the interaction's essential stage, namely the Critical Interval. The ubiquity of this interaction is argued in Section 4. Oblivion is then shown in 5 to underlie several well-known quantum effects. Intervention within the Critical Interval and its unique consequences are studied in 6. Section 7 presents the resolution of the momentum conservation problem. 8 outlines some applications. Section 9 summarizes the work.

# 2. A NON-RECIPROCAL INTERACTION

Let an electron and a positron, with spin states $|z_+\rangle = \frac{1}{\sqrt{2}}(|x_+\rangle + |x_-\rangle)$ and momenta $(p_x)_{e^-} < (p_x)_{e^+}$, $(p_y)_{e^-} = (p_y)_{e^+}$, enter two Stern-Gerlach magnets (drawn for simplicity as beam-splitters) positioned at $(t_0, x_{e^-}, y_0)$ and $(t_0, x_{e^+}, y_0)$ respectively (Fig. 1). The magnets split the particles' paths according to their spins in the $x$-direction:

$$|\psi_{e^-}\rangle = \frac{1}{\sqrt{2}}(|1_{e^-}\rangle + |2_{e^-}\rangle) \text{ and } |\psi_{e^+}\rangle = \frac{1}{\sqrt{2}}(|3_{e^+}\rangle + |4_{e^+}\rangle). \qquad (1)$$

Suppose that technical care has been taken to ensure that, should the particles turn out to reside in the intersecting paths, they would meet, ending up in annihilation.

Let us follow the time evolution of these two wave-functions plus two nearby detectors $|READY\rangle_1$, $|READY\rangle_2$ set to measure the photon emitted upon pair annihilation, changing their states to $|CLICK\rangle_1$ or $|CLICK\rangle_2$.

Initially, the total wave-function is the separable state:

$$|\psi\rangle = \frac{1}{2}(|1_{e^-}\rangle + |2_{e^-}\rangle)(|3_{e^+}\rangle + |4_{e^+}\rangle)|READY\rangle_1 |READY\rangle_2. \quad (2)$$

Depending on their positions at $t_1$ or $t_2$, the particles may (not) annihilate and consequently (not) release a pair of photons, which would in turn (not) trigger one of the detectors. Let these photons, exhibiting the unique superposition emitted/not emitted, be termed "conditional photons".

At $t_0 \leq t < t_1$, then, the superposition is still unchanged as in Eq. 2. At $t_1 < t < t_2$, if a photon pair is emitted, we know that the system ended up in $|2_{e^-}\rangle|3_{e^+}\rangle|CLICK\rangle_1|READY\rangle_2$. Otherwise,

$$|\psi\rangle = \frac{1}{\sqrt{3}}[(|1_{e^-}\rangle + |2_{e^-}\rangle)|4_{e^+}\rangle + |1_{e^-}\rangle|3_{e^+}\rangle]|READY\rangle_1|READY\rangle_2, \quad (3)$$

which is superposition of an interesting type: one component of it is a definite state, as usual, while the other is a superposition in itself.

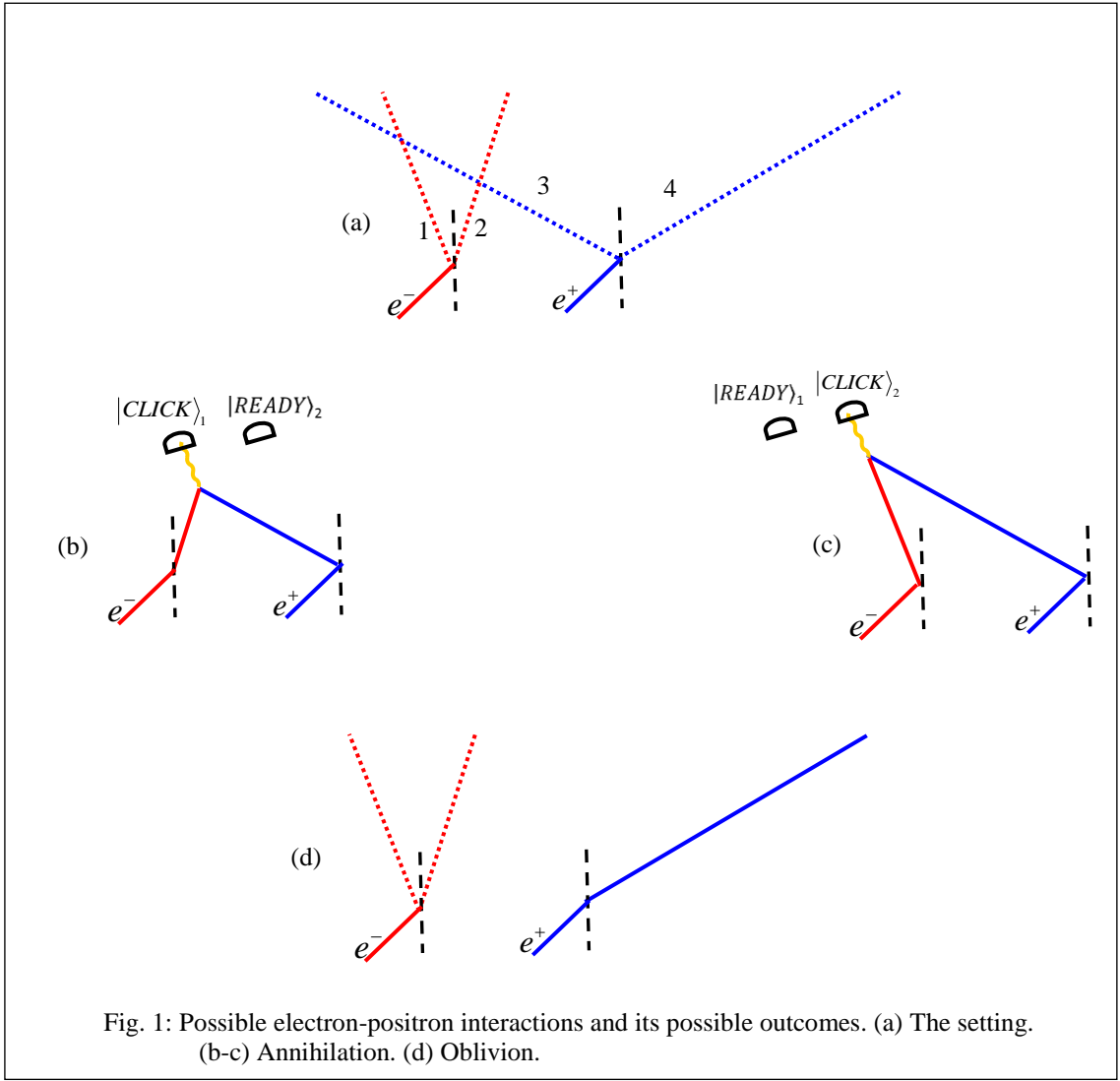

Fig. 1: Possible electron-positron interactions and its possible outcomes. (a) The setting. (b-c) Annihilation. (d) Oblivion.

Similarly at $t > t_2$: If a photon pair is emitted, we know that the particles ended up in paths 1 and 3: $|1_{e^-}\rangle|3_{e^+}\rangle]|READY\rangle_1|CLICK\rangle_2$. Otherwise, we find the entangled state:

$$|\psi\rangle = \frac{1}{\sqrt{2}}(|1_{e^-}\rangle + |2_{e^-}\rangle)|4_{e^+}\rangle|READY\rangle_1|READY\rangle_2, \qquad (4)$$

which is peculiar. The positron is observably affected: If we time-reverse its splitting, it may fail to return to its source.[1] Its momentum has thus changed. Not so with the electron: It remains superposed, hence its time-reversibility remains intact (Fig. 2).

---

[1] This is basically an interference effect, the present setting being the simple Michelson rather than the Mach-Zehnder interferometer.

Summarizing, one party of the interaction "remembers" it through momentum change, while the other remains "oblivious," apparently violating momentum conservation.

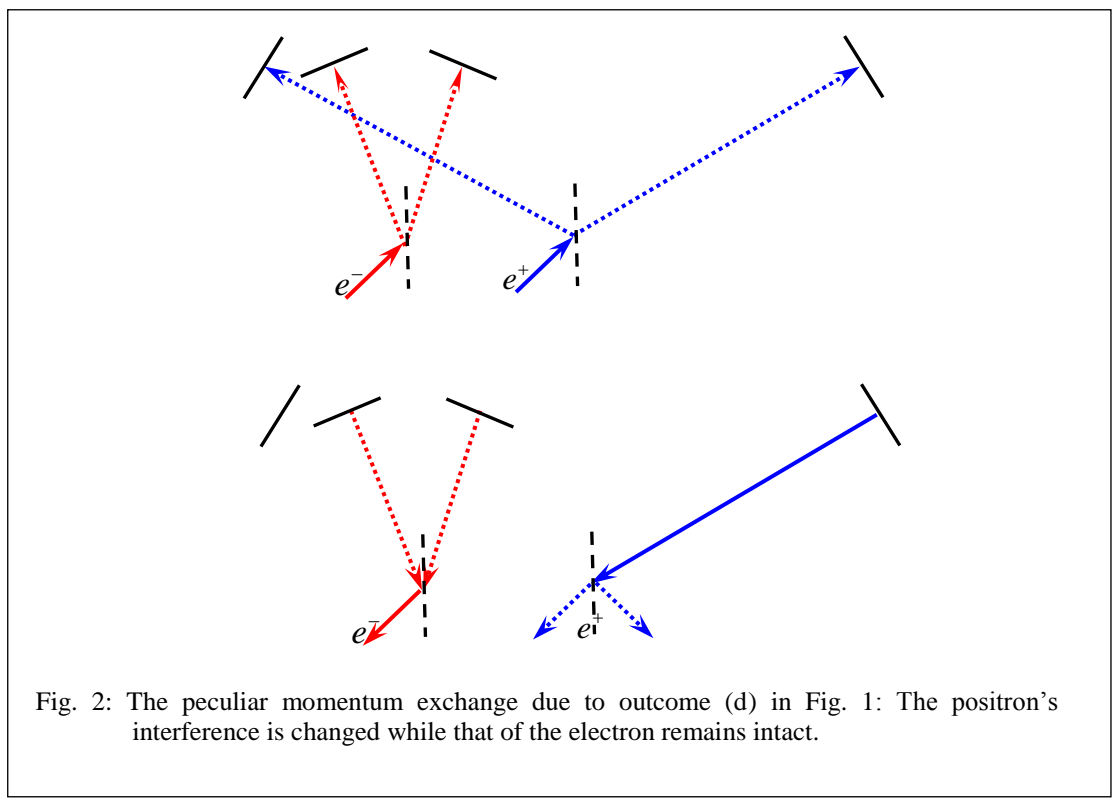

Fig. 2: The peculiar momentum exchange due to outcome (d) in Fig. 1: The positron's interference is changed while that of the electron remains intact.

This is Quantum Oblivion (QO), the varieties of which, the underlying principles and potential applications are studied in what follows.

### 3. THE CRITICAL INTERVAL

It is obviously the intermediate time-interval $t_1 < t < t_2$ that conceals the momentum conservation in QO. The details, however, are no less interesting.

The two particles, during this interval, become partly entangled in both position and momentum. Suppose, *e.g.* that within the interval we reunite the two halves of each particle's wave-function through the original BS, to see whether they return to their source. *Either one* of the particles may fail to do that, on which case the other must *not*. Similarly for their positions. This is *entanglement*, identical to that of the electron-positron pair in Hardy's experiment [2]:

$$|\psi\rangle = \frac{1}{\sqrt{3}}[(|1_{e^-}\rangle + |2_{e^-}\rangle)|4_{e^+}\rangle + |1_{e^-}\rangle|3_{e^+}\rangle]. \qquad (5)$$

The state during this interval is a *higher-order superposition*: Instead of ordinary states such as $|\sigma_z = +1\rangle$ and $|\sigma_z = -1\rangle$, it is composed of a superposed $(|1_{e^-}\rangle + |2_{e^-}\rangle)|4_{e^+}\rangle|I\rangle$ and a non-superposed state $|1_{e^-}\rangle|3_{e^+}\rangle|II\rangle$ as components of the two particles' entanglement.

Equally interesting is the outcome of this partial entanglement, namely, *un*entanglement. This term is proposed instead of the familiar "*dis*entanglement," in order to capture the event's uniqueness. Whereas disentanglement evolves as a direct consequence of entanglement, *e.g.*, the EPR correlations following spin measurements, unentanglement is a process that gives the deceptive impression that entanglement *never took place*. Rephrased in more familiar terms, the wave-function undergoes momentary decoherence followed by "*re*coherence." Notice that, in contrast to the familiar decoherence induced by the macroscopic environment, which is usually believed to be irreversible, here it is momentarily created by the interaction's mere *potential* to become macroscopic.

### 4. THE EFFECT'S UBIQUITY: EVERY DETECTOR'S POINTER MUST BE SUPERPOSED IN THE CONJUGATE VARIABLE

We now submit our main argument. Rather than a curious effect of a specific interaction, *Oblivion is part and parcel of every routine quantum measurement*. Its elucidation can therefore shed new light on the nature of measurement as well as enable some novel varieties of it.

Ordinary quantum measurement requires a basic preparation often considered trivial. Consider *e.g.*, a particle undergoing position measurement (also employed during standard spin measurements [6]). The detector's pointer, positioned at a specific location, reveals the particle's presence in that location by receiving momentum from it. This requires, by definition, that the pointer will have rather precise momentum (preferably 0). In return, the pointer's position must be highly *uncertain*.

Let this tradeoff be illustrated by our first experiment (Fig. 1) plus one modification. In the original version the experiment's two possible interactions are annihilations, which are mutually exclusive. Let us replace annihilation by mere collision (Fig. 3): Simply let two superposed atoms *A1* and *A2* interact exactly like the electron and positron in Fig. 1. Instead of annihilating, then, they merely collide, which can now happen on *both* possible occasions at $t_1$ and $t_2$, namely the two locations where A2 can reside. We know that the two atoms have collided, but remain oblivious about this collision's location.| What we now measure is ordinary momentum exchange, only under a much finer resolution than ordinary quantum measurement, validating our assertion: *Whether atom 2 receives momentum from atom 1 or not, it remains superposed* (Fig. 3).

During the Critical Interval, however, this superposition was affected and restored. The entanglement between the atoms has been more complex than, because both atoms have assumed new possible locations

$$|\psi\rangle = \frac{1}{2}[|1'\rangle|3''\rangle + |2'\rangle|3'\rangle + (|1'\rangle + |2'\rangle)|4\rangle], \qquad (6)$$

which remain undistinguished initial the macroscopic detector that finalizes the interaction (see Sec. 7.1) seals the oblivion.7.2

The generalization is natural. During *every* quantum measurement, the detector's pointer interacts with a particle in the same asymmetric manner as atoms 1 and 2: one wave-function's *part* interacts with the other wave-function's *whole*. To make the analogy complete, recall that in reality the pointer's superposition is continuous rather than discrete. As the pointer thus resides over a wide array of locations, momentum measurement becomes much more precise. This passage from discrete to continuous superposition also opens the door for several interesting interventions, studied in Section 6 below.

**5. OBLIVION'S UNDERLYING OTHER KNOWN QUANTUM EFFECTS**

Not only is Quantum Oblivion essential for every quantum measurement, it is also present in several well-known variants thereof. In what follows we review the most notable examples.

### 5.1 IFM

Interaction-Free Measurement [1] is an intriguing aspect of quantum measurement in that the *non*-occurrence of an event that *could* have occurred gives rise to observable consequences. The analysis of standard measurement in Sec. 4 leads to the straightforward conclusion that Oblivion is IFM's essential stage, occurring during the *($t_1$,$t_2$)* Critical Interval.

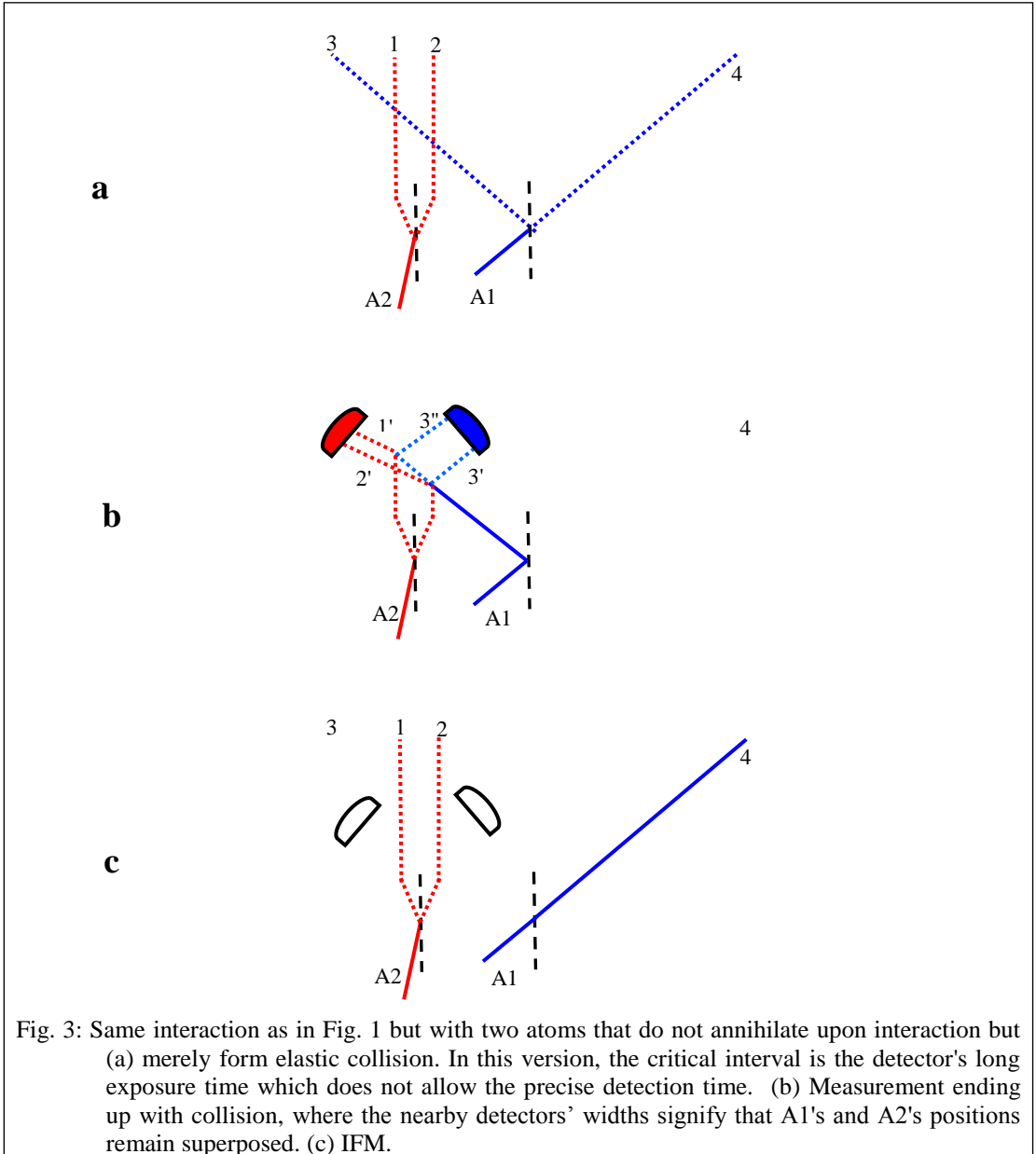

Fig. 3: Same interaction as in Fig. 1 but with two atoms that do not annihilate upon interaction but (a) merely form elastic collision. In this version, the critical interval is the detector's long exposure time which does not allow the precise detection time. (b) Measurement ending up with collision, where the nearby detectors' widths signify that A1's and A2's positions remain superposed. (c) IFM.

Consider again the interaction between the atoms in Fig. 3c, ending up in IFM. Should we intervene during the Critical Interval, measuring whether A2 has gained additional momentum (ignoring its negligible initial own), this would amount to A2's additional position measurement. This measurement would give either *i*) A2 on 1' and A1 on 3", or *ii*) A2 remaining superposed over 1 and 2 while A1 "collapses" to path 4. In other words, entanglement has been formed.

This rapid intervention is not only hard to perform but deliberately avoided in standard quantum measurement, as its outcome mixes position and momentum. If, however, it is performed, some of the resulting effects are very interesting, and studied below.

Ordinary measurement, then, is completed after the Critical Interval, hence Oblivion is *always* part of it: Whether the pointer eventually indicates to have received momentum or not, it first *does* form a mixed position-momentum entanglement with the measured particle, subsequently either *i*) completed into a pure momentum gain or *ii*) obliterated with no trace, namely IFM.

### 5.2    IFM Variants

IFM has inspired the discovery of even more intriguing interactions. As Elitzur and Dolev [3] suggested, IFM can be enhanced by allowing the proverbial "bomb" in the original bomb-testing device to be also superposed. In these "mutual IFMs" the results look even more paradoxical, indicating apparently impossible interactions or conflicting histories.

Hardy's paradox [2] involves a particle and antiparticle symmetrically interacting such that annihilation may or may not occur. If it does not occur, the two particles' wave-functions undergo a change that, when viewed from two reference frames, yields two mutually-exclusive interactions. In fact, this paradox give the major inspiration for our present study of Quantum Oblivion: Our non-reciprocal interaction (Sec. 2) occurs just by allowing one of the particles to intersect two rather than one of the other's paths.

The Quantum Liar paradox [3] involves two atoms that become entangled due to a single photon admitted by only one of them. Because this photon is detected by one detector, making it forever impossible to tell which atom has emitted it, this very ignorance suffices to entangle the two possible origins. Now by Bell's theorem, every atom's emission or non-emission of the photon is nonlocally determined by the measurement chosen to be performed on the other atom. But then, the atom that did not emit the photon could not have taken part in the interaction, hence could not have been entangled in the first place!

These variants of IFM are amenable to the same explanation based on Quantum Oblivion. A Critical Interval has occurred during which entanglement between the particles was formed and then obliterated without a trace.

### 5.3 Aharonov-Bohm Effect

Upon a similar analysis, the Aharonov-Bohm effect [4] turns out to be another distinguished member of the Oblivion-based interactions.

Elaborating on the recent analysis of Vaidman [7], let an electron encircle two concentric cylinders with opposite charges and angular velocities. Together, the two cylinders create a field-free magnetic potential along the electron's trajectory. When entering one arm of the circle, the electron changes the magnetic flux and causes an electromotive force on the cylinders, thereby changing their angular velocity. This way, the electron and cylinders become slightly entangled. Upon leaving the circular trajectory, however, the electron imparts an opposite electromotive force, thereby becoming unentangled with the solenoids (for "unentangled" *vs.* "disentangled" see Sec. 0). As a result, the interaction between the electron and solenoids is "obliterated" – the solenoids return to their initial state, yet the relative AB phase changes are accumulated. The resolution offered above is similar to that offered in Sec. 7.2 for Oblivion in general.

This analogy is not complete only because AB is usually studied as an ordinary quantum-classical measurement, namely with the electron superposed while the flux is macroscopic and thus has definite momentum. Oblivion, on the other hand, can be seen most clearly in quantum-quantum interactions. However, as the analysis of the mirror-detector in Sec. 4 shows, no macroscopic device is free of quantum superposition, hence Oblivion's ubiquity.

### 5.4 Incomplete Measurements: Partial and Weak

Quantum Oblivion is also the source for the paradoxical advantages of some imprecise quantum measurements over the standard one. In these techniques, the critical interval is not completed. This intervention enables the particle and measuring apparatus to take part in an interaction that leaves some intriguing outcomes.

Partial measurement (PM) [8][9] is basically partial IFM, where the portion of the wave-function measured is smaller than that needed for "collapse." Consider a photon traversing an MZI, measured for "which path." To make the measurement partial, the detector should interact not with the entire 50% of the wave-function going on that side, but with any smaller portion of it, such that the measurement's outcome, when not ending with a click, does not indicate with certainty that the photon is *not* on the right, hence it cannot ascertain that the photon *is* on the left. Rather, the photon has increasingly higher probability to be on the left with every incremental "non-detection" on the right.

This imprecision enables observing some phenomena inaccessible to ordinary measurement [8][9]. *Inter alia*, PM can perform complete erasure of its outcome, even when it is arbitrarily close to certainty [8][9]. Consider, *e.g.*, a PM in one arm of a Stern-Gerlach magnet, yielding the particle's spin 99%↑. Rarely, an identical PM on the other Stern-Gerlach arm would yield the same IFM degree, as a result of which the wave-function would return to its initial superposition, totally oblivious of its earlier near-certainty spin-↑ state. Moreover, this erasure, just like positive measurement, exerts nonlocal effect in the EPR experiment. Comfortably, the critical interval in this setting can be made long as one desires. Several realizations of this experiment have been made, see *e.g.* [10][11].

Similarly vital is the role of uncertainty in performing Weak Measurements (WM) [5][12][13]. In this setting, weakness is related to the signal-to-noise ratio, "signal" being the coupling strength between the particle and measuring pointer and "noise" the uncertainty of the pointer's wave-function. By virtue of the pointer's uncertainty, collapse is avoided and the measured system hardly change its state. The meager information thus obtained is obscured by the pointer's noise. Hence, the particle maintains its superposition. This however comes with a great gain: When a sufficiently large ensemble of particles is weakly measured, the signals add up while the noise effectively cancels out.

This technique proved to be an inexhaustible source of novel quantum phenomena and effects hitherto invisible to ordinary quantum measurement [13]. For possible combined WM-PM settings and their predicted effects see [9].

In terms of the Oblivion effect, WM is unique in that the position uncertainty is deliberately made small, followed by strong measurement of the pointer. Consequently the momentum measurement becomes inseparably mixed with position measurement. This is the weak equivalent to the strong intervention within the critical interval – a variant of quantum measurement we study next.

### 5.5 Outcome Manipulations: Quantum Zeno Effect and Quantum Erasure

Quantum Oblivion lies also at the basis of two ingenious techniques for pushing the measurement's outcome towards a desired value, being of great practical and theoretical interest.

*The Quantum Zeno effect*

When oblivion is manipulated to inflict only one undesired outcome of the quantum measurement, a repeated sequence of such actions can safely lead the measurement towards the desired goal. This is the Quantum Zeno Effect [14].

Following [15], consider a photon free to move between two mirrors. Between the mirrors an equidistant beam-splitter is placed, with a very small transmission coefficient $\sin^2\alpha$, where $\alpha \ll 1$. The dynamics is

$$|L\rangle \rightarrow \cos\alpha |L\rangle + \sin\alpha |R\rangle$$
$$|R\rangle \rightarrow -\sin\alpha |L\rangle + \cos\alpha |R\rangle \qquad (7)$$

After $n$ cycles, the state of the particle would be $\cos(n\alpha)|L\rangle + \sin(n\alpha)|R\rangle$, i.e., the particle starts at the left side, but when $n\alpha \approx \frac{\pi}{2}$ it passes to the right side. Let us denote by $n' = \frac{\pi}{2\alpha}$ the critical number of cycles.

Now, if there is a detector on the right side, each cycle is likely to end up in IFM due to the very low explosion probability, and after $n'$ cycles there is still high probability $\cos^{n'}\alpha \simeq (1 - \frac{\alpha^2}{2})^{n'} \simeq 1 - \frac{\pi\alpha}{4}$ of finding the particle on the left. The detector seems to "forget" all the "barely-missed explosions" and thus the particle can be found on the left side where it started. This forgetfulness is another manifestation of QO, hence tampering with it during the CI can produce interesting variants, two of which are presented in the next section.

*Quantum Erasure*

By the above classification, Quantum erasure has a unique place: It is a technique aimed at bringing Oblivion about. It was first suggested by Scully in [16]. "Which path" information encoded in atomic states $|b\rangle$ and $|c\rangle$ in the middle of a measurement, can be erased by a $\pi$-pulse taking the state $|b\rangle$ to $|b'\rangle$ which shortly decays to $|c\rangle$ and by using a common photodetector to collect the photons emitted from the atomic transitions. Loss of "which path" information leads to the re-appearance of interference fringes, even in the delayed-choice version [17].

Erasure of Partial measurement (Sec. 5.4) is complementary to the standard method: In the latter, successful erasure requires that the value erased remains absolutely unknown in any possible record (which, by strict positivist standards, may cast doubt whether anything has been erased in the first place). PM, in contrast, enables perfectly knowing the partial value that has been erased. The price is that erasure's success goes down the closer the partial value gets to 1.

## 6. INTERVENING WITHIN THE CRITICAL INTERVAL: OBLIVION EFFECTS FURTHER STRAINING CLASSICAL NOTIONS

If Oblivion is integral to every standard quantum measurement, remaining unobserved only by the Critical Interval left undisturbed, what happens when this interval *is* disturbed? Subtler ingredients of the measurement process then emerge, their manipulation giving rise to some peculiar effects. Following are some examples.

### 6.1 The Ghostly Mirror

Within our original setup we prepare the electron in the superposition of right and left, but instead of a positron, we prepare a "mirror" polarizable in the *z*-up direction:

$$|\psi(t=0)\rangle_{tot} = \frac{1}{\sqrt{2}}(|Z+\rangle_M (|L_{e^-}\rangle + |R_{e^-}\rangle). \tag{8}$$

Next, we perform a "Hardy split" [2] of the mirror along the *x* direction:

$$|\psi(t=t_0)\rangle_{tot} = \frac{1}{2}(|X+\rangle_M + |X-\rangle_M)(|L_{e^-}\rangle + |R_{e^-}\rangle). \tag{9}$$

In case that no scattering of the electron is recorded in the vicinity of the mirror's front position, the total state turns into:

$$|\psi(t_1 < t < t_2)\rangle_{tot} = \frac{1}{\sqrt{3}}[|X+\rangle_M (|L_{e^-}\rangle + |R_{e^-}\rangle) + |X-\rangle_M |R_{e^-}\rangle]. \tag{10}$$

Now comes the twist. We interfere with the subsequent dynamics within the Critical Interval and measure the mirror's spin along the *z* direction:

$$|\psi_{meas}\rangle_{tot} = \frac{1}{\sqrt{6}}[|Z+\rangle_M (|L_{e^-}\rangle + 2|R_{e^-}\rangle) + |Z-\rangle_M |L_{e^-}\rangle]. \tag{11}$$

In case the outcome is spin down:

$$|\psi_{proj}\rangle_{tot} = |Z-\rangle_M |L_{e^-}\rangle. \tag{12}$$

We have a non-zero probability to end with the photon on the *Left* and mirror on the *Right*.

### 6.2 Nonlocality in the Dicke IFM

Following is a variant of Dicke's paradox [18]. Consider (as in Fig. 4.a) a wave-function of a particle widely superposed in the following spatial manner. A large fraction of the wave-function's amplitude $\sqrt{1-\varepsilon^2}$ is continuously spread over a wide region $L$ ("tray") while the remaining $\varepsilon$, located far away, is homogenously spread over a much narrower region $\ell$ ("spoon"). Let both tray and spoon be surrounded with a sufficient array of close detectors, such that if the particle turns out to reside in one of them, and moreover receives some momentum which makes it bounce, it will be detected by one of the detectors.

Now suppose we make a position measurement on the tray, and consider the very rare case where this measurement turns out to be IFM, indicating that the particle is rather in the faraway spoon (Fig. 4.b). This extreme decrease in the particle's position uncertainty must take its toll on its momentum uncertainty, previously very small. Consequently, a detector positioned next to the spoon may suddenly be kicked by the particle, which was so far well behaved with respect to its momentum (while barely being there probabilistically), now going wild as the probability for jumping out goes to 1.

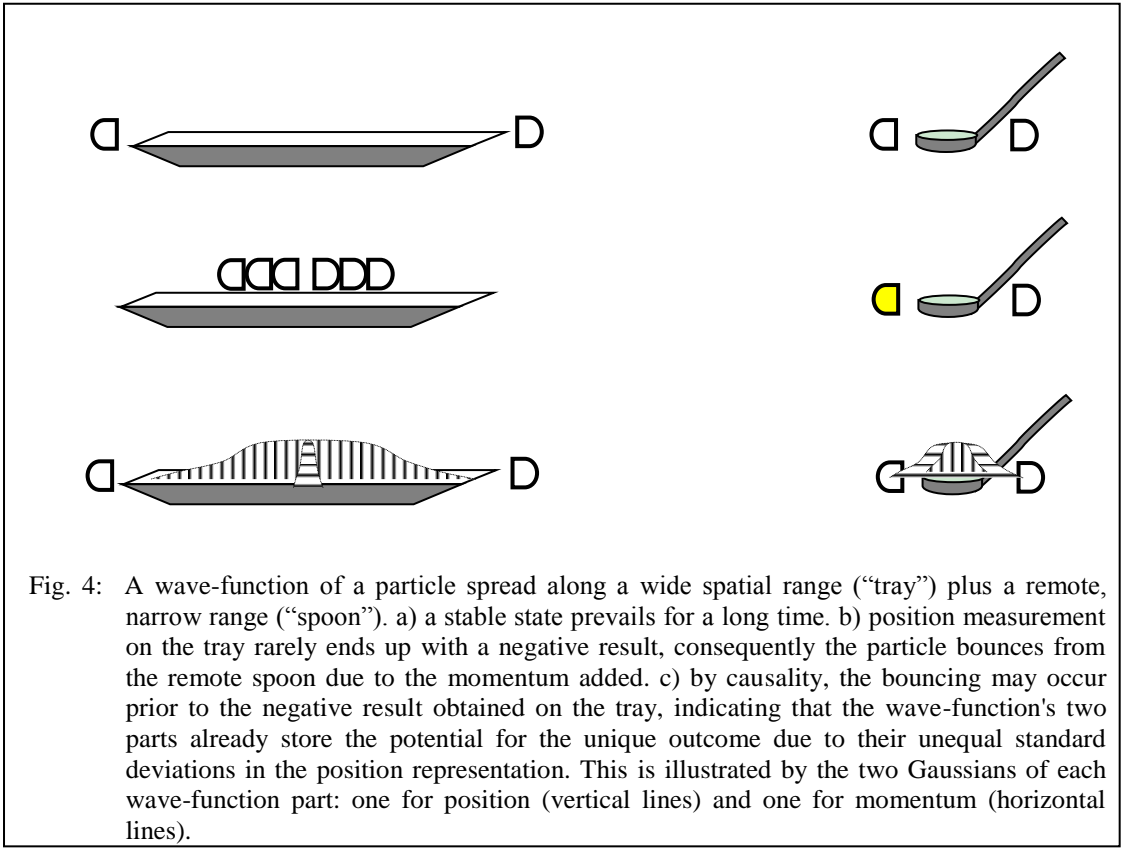

Fig. 4: A wave-function of a particle spread along a wide spatial range ("tray") plus a remote, narrow range ("spoon"). a) a stable state prevails for a long time. b) position measurement on the tray rarely ends up with a negative result, consequently the particle bounces from the remote spoon due to the momentum added. c) by causality, the bouncing may occur prior to the negative result obtained on the tray, indicating that the wave-function's two parts already store the potential for the unique outcome due to their unequal standard deviations in the position representation. This is illustrated by the two Gaussians of each wave-function part: one for position (vertical lines) and one for momentum (horizontal lines).

True, this case is extremely rare, and to make the orders of magnitude even approach realistic scales the superposition must be extremely unique. Moreover, causality obliges an equal probability for a spontaneous jump of the particle from the spoon even before the tray was measured. Here, the detector next to the spoon performs position-plus-momentum measurement of that wave-function's portion. But in this case another odd quantum process is revealed. The original superposition is a complex one. This means that the detector, although placed too far from the spoon, still interacts with the Gaussian "tail" of the particle's widely superposed momentum, surrounding tiny the spoon.

$$\begin{aligned}\psi(t) &= \sqrt{1-\varepsilon^2}\, N_1 \exp[-(x-x_1)^2/2L^2] + N_2 \varepsilon \exp[-(x-x_2)^2/2\ell^2] \\ &\Rightarrow N_f \exp[-(x-x_2)^2/2\ell^2] \\ &\Rightarrow Click \end{aligned} \qquad (13)$$

where the tray/spoon are located at $x_1$ / $x_2$ and the $N$s are normalization factors.

We believe this variant of quantum nonlocality merits a closer study.

### 6.3 Two variations of the the Quantum Zeno Effect (QZE)

Within the CI, The Quantum Zeno Effect (Sec. 5.5) can also undergo new twists:

*Variation 1 (counterfactual measurement):* Assuming now that the bomb is quantum and has spin-1/2, we encounter a surprising effect: Suppose that the bomb is prepared with $|\sigma_x = +1\rangle$, but reflection of the photon is possible only in the cases where the spin of the bomb is $|\sigma_z = +1\rangle$. If we now find the photon on the left after *n'* cycles, we are likely to deduce that the bomb has $|\sigma_z = +1\rangle$. Hence, the photon determined the spin of the bomb, without even reaching it!

*Variation 2 (Ghostly entanglement):* Divide now the box into 3 parts, in a similar manner to the above case, where two quantum bombs are located at the two sides, and the photon starts from the middle. After *n'* cycles we look for the photon on the middle part. The meaning of not finding it is *interaction-free-entanglement*: The photon singled out the product state: $|\sigma_z = +1\rangle|\sigma_z = +1\rangle$ of the mirrors leaving them entangled.

### 7. CONCEPTUAL ISSUES

Out of many fascinating questions emerging from the above study, let us deal with the most urging ones.

### 7.1 The unnoticed role of "conditional photons"

The above analysis dealt with our more robust version of the oblivion experiment, namely the one involving atoms rather than a particle-antiparticle pair. There is, however, a very interesting element in the particle-antiparticle version, namely, the photons (not) created upon the (non-)occurrence of annihilation. To stress their nature, suppose that the detectors set to detect them are very far away.

Here too, the interaction is finalized by the two detectors *not* indicating that they have absorbed a photon – for the simple reason that there was no photon to absorb. What is unique in these photons, then, is their superposition: Not "left"/"right," as in the ordinary MZI measurement,

but rather "existent"/"inexistent."

We term these entities "conditional particles," and they will prove essential for the momentum conservation issue below. We also submit that further study may reveal important insights about the role such conditional particles play in other interactions.

### 7.2 Is momentum conserved?

Momentum is conserved in QO for the simple reason that, although the interaction in question takes place between two quantum particles, a macroscopic detector finalizes the process. We now understand that to our time-reversal in Sec. 2, we should have added also a pair of mirrors to reflect back the conditional photons as we did with the particles (Fig. 1), in which case time-reversal of the entire process would perfectly succeed. It is therefore the use of macroscopic detectors, for either the conditional photons or the particles themselves, that introduced both failure of reversibility and momentum conservation problems. The interaction described in Sect. 24 (Fig. 3), namely the momentum exchange between two atoms, is therefore as follows:

1. Initially, A1 and A2 have position uncertainties, while their momenta are almost certain.
2. During the Critical Interval $(t_1, t_2)$ *i.e.*, the time needed by the two detectors, under low time-resolution, to detect the atoms' collision, each wave-function's half interacts with the other's half, such that the two particles form a symmetric entanglement of their positions $|\psi\rangle = \frac{1}{\sqrt{2}}(|1'\rangle|3''\rangle + |2'\rangle|3'\rangle)$ which make their momenta similarly entangled.
3. After the Critical Interval, A1's half wave-function has interacted with A2's entire wave-function, forming an asymmetric interaction.
4. Then, as A1 and A2 interact with the two detector's pointers, they enter together a new, fourfold entanglement.
5. A Critical Interval similar to that of (1)-(2) now occurs with the two pointers. Thus the pointers' momenta become also uncertain for an even briefer Critical Interval. However, because these pointers are attached to a macroscopic amplifying mechanism, they no longer allow time-reversing the process.

6. After the CI, as the two pointers' momenta become certain again, their precise momentum exchanges with A1 and A2 become forever unknown.

In other words, the oblivious particle, which has exchanged momentum with the other particle during the Critical Interval, has used that other particle to nonlocally transfer its momentum to the *third and fourth parties* joining the interaction, namely the two macroscopic pointers, setting the future boundary condition of the experiment. Momentum is transferred together with, so to speak, the entire "memory" of the interaction between the atoms.

### 7.3  IFM Demystified

For one of us (AE), this work signifies coming full circle, as some 20 years ago he has been privileged to find out, together with Vaidman [1], some new effects of Interaction-Free Measurement. Let us therefore make a comment about this effect's interpretation.

How can a photon know whether a bomb is capable of exploding or not, without bringing this explosion? It is this "if" aspect of the interaction which gives it its uncanny impression, understandably eliciting interpretations along the most extreme schools in quantum philosophy. Consider the following dominance ones:

The Many-Worlds Interpretation: This was Vaidman's own choice. It is in a parallel world that the bomb's explosion takes place, affecting our world through the interference which rarely enables two such histories reunite again.

The Copenhagen Interpretation: It famously treats quantum mechanics as a theory dealing only with the observer's knowledge, refusing to make any ontological statement about things really happening out there. Zeilinger [19] is an eloquent proponent of a modern variety of this school, which emphasizes information as physics' sole subject matter. In the case of IFM, the Copenhagenist's question is simple: Does the observer gain new information following the detector's *silence*? A positive answer logically obliges that, in compliance with the uncertainty principle alone, uncertainty must plague the photon's momentum, and since information is the only relevant currency, no wonder that a photon, supposedly "out there," complies with this obligation.

We submit that the analysis of quantum oblivion brings this question from metaphysics back to laboratory physics. Instead of addressing only information, we can inquire the physical process that this information describes. It then turns out that interaction-free measurement is not really free. Rather, the parties involved have been momentarily subject to entanglement which can be physically detected, and only later liberated themselves through oblivion.

### 7.4  Is Quantum Measurement Explained?

Having demystified IFM, we wish to stress that quantum measurement itself has *not* been demystified by this analysis, and still manifests the famous spatial and temporal oddities, as sorely missing a satisfactory explanation as it has been for the last decades. We do believe, however, that further elaboration of quantum oblivion may shed new light on the more general issue as well.

We refer to some time-symmetric formulations of quantum mechanics, and especially, the Two-State-Vector Formalism (TSVF) [20][21][22]. One unique feature of this interpretation is the derivation of positive-plus-negative weak values exchanges, which are created by the complimentary wave-function coming back from the future absorber to the past source (see for instance [23]).

In our present work, still underway, we study the possibility that such interactions are part and parcel of *every* quantum interaction, as already suggested by the possibility of negative energy exchanges implied in Wheeler and Feynman's "absorber theory," Dirac's hypothesis of "holes" in the vacuum "particles' ocean" that behave like particles with negative mass, and Cramer's Transactional Interpretation. Novel predictions based on this model may enable testing it.

## 8. SUMMARY

Since the days of Bohr and Einstein, quantum paradoxes are famously instrumental in advancing our understanding of the underlying theory. We believe this is the case with the present paradox as well elucidating some aspects of uncertainty and entanglement. The absence of reciprocity in the asymmetric interaction described in Section 2 introduces the ubiquitous Oblivion which lies at the heart of IFM, Hardy's paradox, partial and weak measurements, AB other effects. Sending a known variable into the sea of uncertainty may end up in its drowning there, disappearing from memory – not only the observer's but the memory of Nature herself.

In the quantum world, however, it often turns out that "ignorance is power": subjecting a variable to quantum uncertainty endows it also with compensating novel properties. In our case, leaving the Critical Interval unobserved gives rise to entanglement in some cases, while saving us from collapse in others. A few applications combining this analysis with the QZE were also presented, but more are yet to come [24].

## Acknowledgements

It is a pleasure to thank Yakir Aharonov and Anton Zeilinger for helpful comments and discussions. This work has been supported in part by the Israel Science Foundation Grant No. 1311/14. We are also thankful to Templeton Foundation and Robert Boyd for a very fruitful discussion and exchange of ideas.

# References

[1] A.C. Elitzur, L. Vaidman, *Found. Phys.* **23** (1993) 987–997.

[2] L. Hardy, *Phys. Rev. Lett.* **68** (1992) 2981–2984.

[3] A.C. Elitzur, S. Dolev, *AIP Conf. Proc.* **863** - *Frontiers of Time: Retrocausation – Experiment and Theory* (2006) 27-44.

[4] Y. Aharonov, D. Bohm, *Phys. Rev.* **115**, (1959) 485-491.

[5] Y. Aharonov, D. Albert and L. Vaidman, *Phys. Rev. Lett.* **60** (1988) 1351-1354.

[6] K.K Wan, R.G. McLean, *J. Phys. A: Math. Gen.* **24** (1991) L425.

[7] L. Vaidman, *Phys. Rev. A* **86** (2012) 040101.

[8] A.C. Elitzur, S. Dolev, *Phys. Rev. A* **63** (2001) 062109.

[9] A.C Elitzur, E. Cohen, *AIP Conf. Proc.* ***1408***: *Quantum Retrocausation: Theory and Experiment* (2011) 120-131.

[10] X.Y. Xu, J.S. Xu, C. F. Li, Y. Zou, G.C. Guo. *Phys. Rev. A* **83** (2011) 010101.

[11] M. Jakob, J. Bergou, *Phys. Rev. A* **66** (2002) 062107.

[12] B. Tamir, E. Cohen, *Quanta* **2** (2013) 7-17.

[13] Y. Aharonov, E. Cohen, A.C. Elitzur, *Phys. Rev. A.* **89** (2014) 052105.

[14] E.C.G. Sudarshan, B. Misra, *J. Math. Phys.* **18** (1977) 756–763.

[15] L. Vaidman (2014) arXiv preprint, arXiv:1410.2723.

[16] M. Scully and K. Drühl, *Phy. Rev. A* **25** (1982) 2208.

[17] Y. Kim, R. Yu, S. Kulik, Y. Shih, M.O. Scully, *Phy. Rev. Lett.* **84** (2000) 1.

[18] R.H. Dicke, *Am. J. Phys.* **49** (1981) 925-930.

[19] J. Kofler, A. Zeilinger (2006), http://www.mpq.mpg.de/.

[20] Y. Aharonov, P.G. Bergman, J.L. Lebowitz, *Phys. Rev.* **134** (1964) 1410-1416.

[21] Y. Aharonov, L. Vaidman, *Lect. Notes Phys.* **734** (2008) 399-447.

[22] Y. Aharonov, E. Cohen, E. Gruss, T. Landsberger, *Quantum Stud.: Math. Found.* **1** (2014).

[23] Y. Aharonov, A. Botero, S. Popescu, B. Reznik, J. Tollaksen, *Phys. Lett. A* **301** (2002) 130-138.

[24] A.C. Elitzur, E. Cohen, The quantum undoing hypothesis: negative momentum exchanges as the hidden ingredient in null quantum measurements – Forthcoming.